\documentclass{aa}
\usepackage{graphicx}
\begin{document}
   \title{INTEGRAL\thanks{INTEGRAL is an ESA project with instruments and science data centre funded by ESA member states (especally the PI countries: Denmark, France, Germany, Italy, Switzerland, Spain), Czech Republic and Poland, and with the partecipation of Russia and the USA.} high energy behaviour of 4U 1812-12}
    
%  \subtitle{}

   \author{A. Tarana \inst{1}
             \and A. Bazzano \inst{1}
             \and P. Ubertini\inst{1}
             \and M. Cocchi\inst{1}
             \and D. G\"{o}tz\inst{2}
             \and F. Capitanio\inst{1}$^{,}$\inst{3}
             \and A. J. Bird\inst{3}
             \and M. Fiocchi\inst{1}
          %\inst{1}
              %\fnmsep\thanks%{Just to show the usage
          %of the elements in the author field
          }

%   \offprints{P. S. Negi}
 \institute{Istituto di Astrofisica Spaziale e Fisica Cosmica-INAF,
via del Fosso del Cavaliere 100, 00133 Roma
\and Istituto di Astrofisica Spaziale e Fisica Cosmica-INAF, via  E. Bassini 15, 20133 Milano
\and Department of Physics and Astronomy, Southampton
  University, SO17 1BJ Southampton, U.K.\\
\email{Antonella.Tarana@rm.iasf.cnr.it}
}
\date{Received ; accepted }
   \abstract{The low mass X-ray binary system 4U 1812-12
   was monitored with the INTEGRAL observatory in the period 2003-2004 and 
with BeppoSAX on April 20, 2000. We
   report here on  the spectral and temporal analysis of both persistent and
   burst emission. The full data set confirms the persistent nature of this burster, and reveals the presence
   of emission up to 200 keV. The persistent spectrum is well described by a
   comptonization (CompTT) model plus a soft blackbody component. The source
   was observed in a hard spectral state with a
   1-200 keV luminosity of 2$\times$10$^{36}$ ergs s$^{-1}$
  and L/L$_{Edd}$$\sim$1$\%$ and no meaningful flux variation has been revealed,
  as also confirmed by a 2004 RXTE observation. We have also detected 4 bursts showing double peaked profiles and blackbody spectra
   with temperatures ranging from 1.9 to 3.1 keV.
}
%\titlerunning{4U 1812-12}
\authorrunning{Antonella Tarana}
\titlerunning{INTEGRAL high energy behaviour of 4U 1812-12}
\maketitle
%
%________________________________________________________________

\section{Introduction}
% emissione continua e bursts
4U 1812-12 was discovered as a weak galactic source by the Uhuru satellite in 1970 [\cite{forman}].\\
Type 1 X-ray bursts were detected for the first time with Hakucho and
the burst detection allowed the determination of the nature of the compact
object as a neutron star in a Low Mass X-ray Binary (LMXB) system [\cite{murak}].
During the Galactic Bulge monitoring campaign performed in the period
1996-2001 with the WFCs on board BeppoSAX, several
X-ray bursts were detected and in most
of them a clear  photospheric radius expansion due to Eddington-limited
burst luminosity was observed. This allowed the estimation of the source distance at
4.1$\pm$0.5 kpc. At energies above 5 keV, the bursts showed a double peaked profile that became more evident at higher energy [\cite{cocchi}].\\
A broad band spectrum of the persistent emission and rapid X-ray variability has been revealed during
simultaneous observations with BeppoSAX and RXTE on April 20, 2000. Emission above 20 keV has been reported while the source was in a
hard spectral state with a 1-200 keV flux of 1.1$\times$10$^{-9}$ ergs
s$^{-1}$ cm$^{-2}$ [\cite{barret}].\\
The persistent emission, revealed by Chandra on June 14, 2000, is characterized
by a  2-10 keV flux of about 4$\times$10$^{-10}$ ergs s$^{-1}$ cm$^{-2}$
[\cite{chandra}].
The source was still in the hard state during the RXTE observation in June and July 2001 with a 2-10 keV flux of 3.8$\times$10$^{-10}$ ergs s$^{-1}$ cm$^{-2}$ [\cite{muno}].
\\
Like most of the bursters, 4U 1812-12 is
classified as an atoll source [\cite{atoll}].\\\\
%osservazioni integral
Since March 2003, this source has been observed, and is still being monitored,
by the INTEGRAL observatory [\cite{wink}] in the
framework of the Galactic Centre Deep Exposure programme [\cite{wink2}].\\
In this paper we report on the analysis performed with IBIS/INTEGRAL [\cite{uber}] during the period
2003 March 11$^{th}$ up to  2004 October 20$^{th}$, on both the persistent and bursting behaviour. For the first time a hard
X-ray continuum emission up to 200 keV has been detected, and moreover 4 bursts have been revealed during this long observation period.
%For the continuum emission study, it is essential to have as much as possible
%a broad band spectral data, so we have combined  the IBIS data with the RXTE
%data obtained on April 20 2000. Even if these data are  not
%simultaneous\footnote{The only 4U 1812-12 observation with RXTE in 2004 is not yet in the public archive}, combining spectra is meaningful in view
%of the low observed variability of this source (see section 3 for more details).
%We do not report on JEM-X results, because unluckley 4U 1812-12 was not in the JEM-X instrument [\cite{lund}]
%field of view during most of the obsevations and also because the effective exposure
%is small due to the narrow field of view of the instrument in comparison to
%IBIS and the dithering strategy of the INTEGRAL observation.
\section{Observations and Data Analysis}
% su integral
The IBIS instrument onboard the INTEGRAL satellite is a coded mask imager with a wide field of view (29$^\circ\times$29$^\circ$ at zero
response and 9$^\circ\times$9$^\circ$ fully coded). IBIS is composed of two
detector layers: ISGRI (15 keV - 1 MeV ) [\cite{leb}] and PICsIT (175 keV - 10
MeV) [\cite{labanti}].
%and a Partially Coded Field of View of 19$^\circ$.
 An INTEGRAL observation consists of a group of pointings, called science
 windows (scws) each lasting about 2000 seconds [\cite{win3}].
% sul set di dati
We analysed all the data in which 4U 1812-12 was within the ISGRI/IBIS detector field
of view (14$^\circ$ from the field centre).\\
The observations covered a non-continuous period from 2003 March 11$^{th}$ to
2004 October 20$^{th}$ (from orbit 49 to 246) for a total of 1322 scws. We extracted the light curves in four energy bands and computed the 40-60/20-40 keV
hardness ratio for the whole period.\\
For the spectral analysis only, we selected and used a data set with a total
time of about 165 ks when the source was in the fully coded field of view
(within 4.5 degrees from the centre of the field of view). With this data set,
the source intensity determination is not affected  by  possible systematic
errors due to calibration uncertainties in the off-axis response.\\
During the observing compaign, 4 bursts triggered the INTEGRAL Burst Alert
System (IBAS) [\cite{mereg}] from a direction consistent with the 4U 1812-12 error box.
For each burst, we have extracted the light curve with a temporal bin of 1
second allowing analysis of the burst profiles. We have also extracted the burst
spectra in order to compare the spectral parameters of the bursts.\\
The IBIS data were processed using the 4.2 version of the Off-Line Scientific
Analysis (OSA) [\cite{gold}] software,
 released by the INTEGRAL Scientific Data Centre (ISDC)  [\cite{curvosier}].\\
The response matrix used for the spectral analysis is the standard 2048 channels
matrix, logarithmically rebinned to 64 channels.\\
We do not report on JEM-X [\cite{lund}] results, because when 4U 1812-12 was in the JEM-X field of view, it was not detected in the mosaic
image.\\    
%We do not report on JEM-X [\cite{lund}] results, because during most of the
%observations, 4U 1812-12 was unfortunately not in the JEM-X instrument field of view and also because the effective exposure
%is small due to the narrow field of view of the instrument in comparison to
%IBIS and due to the dithering strategy of the INTEGRAL observation.\\
For the continuum emission study, it is essential to have as 
 broad a spectral coverage as possible. For this reason we have combined  the IBIS (20-200 keV) data
with the BeppoSAX LECS (0.5-4 keV) and MECS (1.5-10 keV)
%\cita referenze beppoSAX
%%%%%%%%%%%%%%%%%%prima le cl stavano qui!!!
%
% curve di luce- hardness ratio*********************************************************************************>
%
%\vspace{0.1cm}
\begin{figure*}[ht]
 \centering
\includegraphics[height=18cm,width=16.5cm, angle=90]{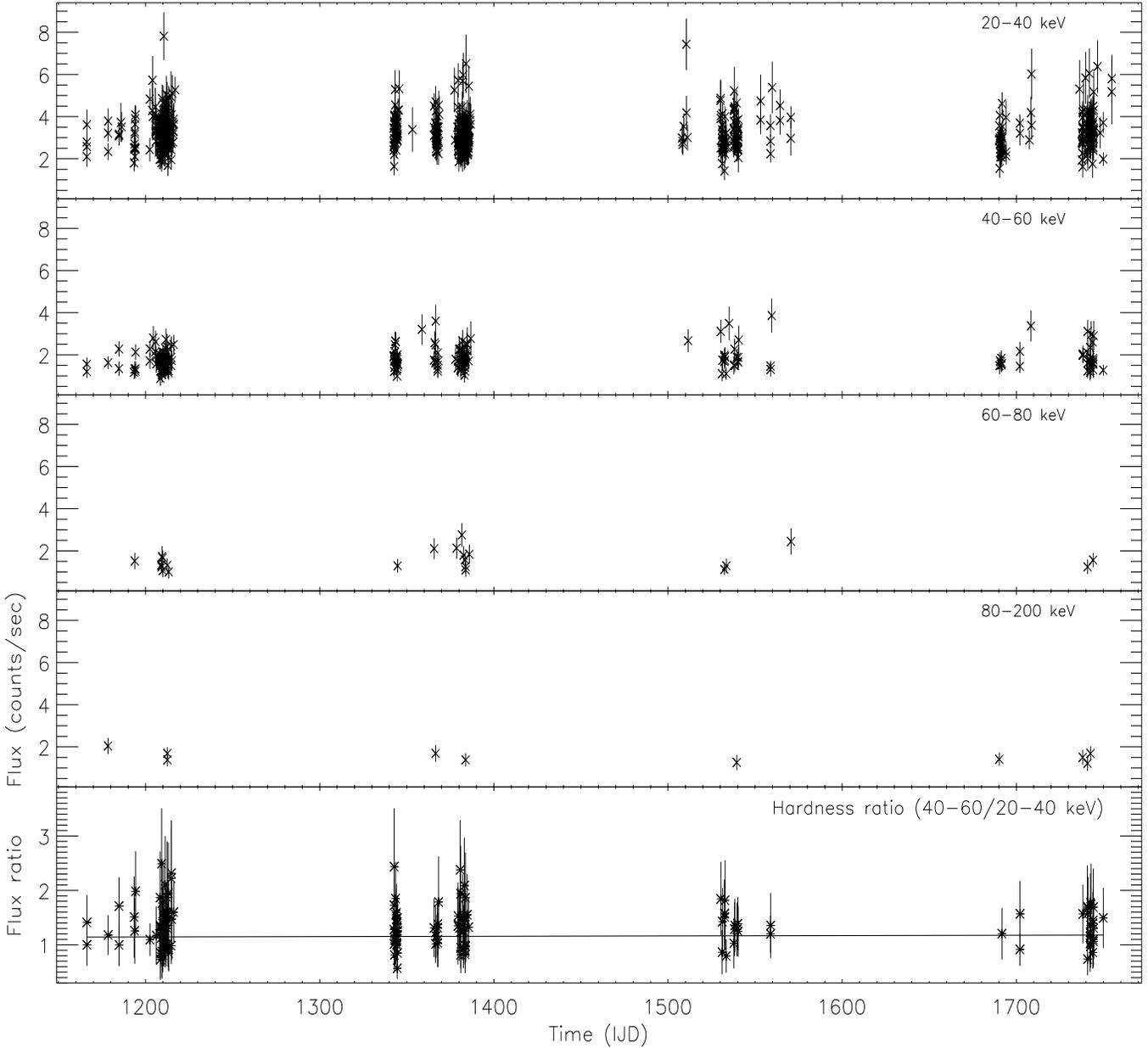}
\caption{4U 1812-12 IBIS/ISGRI light curves, in the  20-40, 40-60, 60-80,
  80-200 keV energy bands; and the hardness ratio of the 40-60 to 20-40 keV flux. The observation period is between 2003 March 11$^{th}$ to 2004 October 20$^{th}$, and each data point rapresents a single pointing.}
\label{cluce}
%%   \includegraphics[height=5cm,width=18cm]{provatitolo.ps}
%%\caption{Hardness ratio of flux 40-60 keV to 20-40 keV.}
%%\label{hr}
 \end{figure*}
%
%%\begin{figure*}[ht]
 %%\centering
% \vspace{0.36cm}
  %% \includegraphics[height=5.3cm,width=18cm]{newHRtot.ps}
%%\caption{Hardness ratio of flux 40-60 keV to 20-40 keV.}
%%\label{hr}
 %%\end{figure*}
%***************************************************************************************************************************<
data obtained in 2000. Even if these data are  not
simultaneous, combining spectra is meaningful in view
of the low observed variability of this source (see section 3 for more details).
 We have examined the public data obtained with BeppoSAX between 20 April at  18:51 UT to 21 April 2000 at 13:23 UT. The publicly available data from the MECS and LECS instruments have been downloaded from the on line archive of BeppoSAX\footnote{http://www.asdc.asi.it/}.\\
The XANADU software package XSPEC (version 11.3.2) has been used for the spectral analysis.
\section{Continuum emission Results}
During the INTEGRAL monitoring of the Galactic Centre, the burster
 4U 1812-12 was detected at position R.A.=18$^h$15$^m$07$^s$ Dec=$-$12$^{\circ}$:05$^{\prime}$:45$^{\prime}$$^{\prime}$  (J2000) with an error
 of 1$^{\prime}$. Figure \ref{cluce} shows the ISGRI/IBIS
light curves in the 20-40, 40-60, 60-80 and 80-200 keV energy bands for
the whole monitoring period. Each point corresponds to the source count rate (at a signal to noise ratio $>$3$\sigma$) for a single pointing. As can be seen, the
source is significantly detected up to 200 keV and shows
the same time behaviour in all energy bands, revealing a persistent and moderately variable continuum emission. In the 20-40 keV
band, the average source flux corresponds to $\sim$ 26 mCrab.\\
To search for a possible correlation between flux
and spectral variation, we derived the hardness ratio
in the 40-60$/$20-40 keV flux. The  null slope (with a $\chi^{2}_{red}\sim$ 1) in the bottom panel of Figure \ref{cluce}
%\ref{hr}
indicates that there is no spectral variability.\\
Spectra have been extracted for each pointing and, in view of the lack of flux
variation, it has been possible to add all the spectra to increase the signal to noise
ratio.
%
%+++++++++++++++++++++++++++++++++++++++++++++++++++++++++++++++++++++++++++++++++++++++++++
%
%
% TABELLA con i valori spettrali emissione continua:
%
\begin{table}[ht]
\begin{center}
\caption{Spectral fitting parameters of the 4U 1812-12 persistent emission. The best fit is a sum of comptonization (CompTT) and blackbody (bb) models.}
\label{tabspetot}
 \begin{tabular}{rl}
%\hline
 %\cline{1-2}
 \multicolumn{2}{l}{Spectral Model: wabs$\times$(bb+CompTT)}\\
\hline
\hline
Parameters & value  \\
\hline
N$_{H}$ &1.5$^{+0.1}_{-0.1}$ $\times10^{22}$ cm$^{-2}$\\
$\textit{T}_{0}$(keV)   & 0.3$^{+0.1}_{-0.1}$ \\
\vspace{0.04cm}
         $\textit{kT}_\textit{e}$(keV) & 17.5$^{+1.7}_{-1.4}$ \\
 \vspace{0.04cm}
           $\tau$ & 2.3$^{+0.2}_{-0.2}$ \\
 \vspace{0.04cm}
        norm$_{comptt}$ & 7.6$^{+1.0}_{-0.9}$$\times$10$^{-3}$ \\ 
\vspace{0.04cm}
           $\textit{kT}_{bb}$(keV) & 0.6$^{+0.1}_{-0.1}$   \\
\vspace{0.04cm}
           norm$_{bb}$ & 7.5$^{+1.4}_{-1.4}$$\times$10$^{-4}$\\
\hline
\vspace{0.01cm}
$\chi_{r}^2$(d.o.f) & 1.03 (514)\\
\hline
Flux$_{20-200\hspace{2pt}\textrm{keV}}$ & 4.2$\times$10$^{-10}$erg cm$^{-2}$s$^{-1}$ \\
Flux$_{1-10\hspace{2pt}\textrm{keV}}$ & 3.2$\times$10$^{-10}$erg cm$^{-2}$s$^{-1}$ \\
%CompPS & Parameters                        &      &            \\
%        &  $T0$(keV) =                     &      &     \\
%       & $\textit{kT}_\textit{e}$(keV)=    &      &     \\
%       &     &      &     \\
\hline
\hline
\end{tabular}
\end{center}
%\caption{\small{\emph{IBIS.}}}
%\label{tabspetot}
\end{table}
% FINE TABELLA
%
%++++++++++++++++++++++++++++++++++++++++++++++++++++++++++++++++++++++++++
%
%++++++++++++++++++++++++++++++++++++++++++++++++++++++++++++++++++++++++++
% ___________SPETTRO
\begin{figure}[ht]
 \centering
 \includegraphics[height=8.8cm,width=7cm, angle=-90]{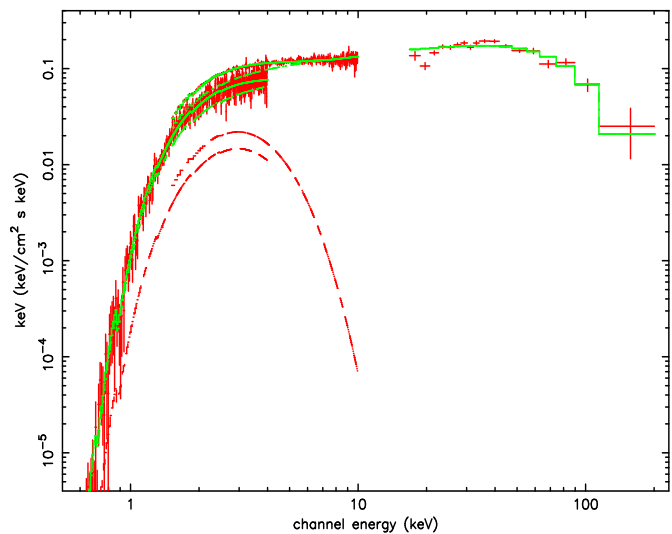}
%\vspace{0.5cm}
%  \includegraphics[height=9cm,width=3cm, angle=-90]{delchi_con2.ps}
\caption{INTEGRAL/IBIS/ISGRI and BeppoSAX/LECS and MECS spectrum of the persistent emission
 of 4U 1812-12, with the CompTT plus blackbody models.}
 \label{spettro}
%\includegraphics[height=9cm,width=3cm, angle=-90]{delchi_senza2.ps}
%\caption{Residuals of the fit without the gaussian component and frozing all the
%  best fit parameters of table \ref{tabspetot}.}
%\label{riga}
 \end{figure}
%________________________________________________________________
%%%%%%%%%%%%%%%%%%%%%%% ecesso riga
%\begin{figure}[ht]
% \centering
%\includegraphics[height=9cm,width=3cm, angle=-90]{delchi_senza2.ps}
%\caption{Residuals of the fit without the gaussian component and with all the
%  best fit parameters of table \ref{tabspetot} frozen.}
%\label{riga}
%\end{figure}
%%%%%%%%%%%%%
The final spectrum has been obtained for a total time of 165 ks, and does
not include any of the burst events.\\
This spectrum is well represented by a thermal comptonization model that
indicates that the source was in the hard spectral state, the same state in which
the source was found during the contemporaneous BeppoSAX and RXTE observation in
2000 [\cite{barret}]. Because of this and  of the weak variability of the
source, we combined the 2000 BeppoSAX low energy data with the IBIS data, even though they are not simultaneous. The intercalibration between the three instruments shows agreement: the relative normalizzation factor of ISGRI with respect to LECS and MECS is 1.03$\pm$0.05, while the relative normalisation factor of  LECS with respect to MECS is 0.67$\pm$0.07. The spectrum thus obtained has been fitted with different models. We always assume a systematic error of 0.01 for
the MECS and LECS and of 0.02 for ISGRI.
The best fit model, a
comptonization model compTT [\cite{sun}] plus a low energy blackbody component, is reported in Table
\ref{tabspetot}, while the spectrum is shown in Figure \ref{spettro}. The low energy data can also be fitted by the multicolor disk black body
(diskbb) model [\cite{mitsuda}], leading to similar parameters and
$\chi_{r}^2$ results.\\
%The electron temparature value is lower ($\textit{kT}_\textit{e}$=15.18$^{+1.7}_{-2.4}$
%keV) than the value given by
%BeppoSax observations ($kT_\textit{e}$=36.0$^{+78.0}_{-9.0}$ keV)[\cite{barret}].\\
%
%
%The combined RXTE and IBIS data result in a better evaluation of the comptonization model parameters if compared with BeppoSAX observation value ($kT_\textit{e}$=36.0$^{+78.0}_{-9.0}$ keV,
%$\tau$=3.0$^{+0.7}_{-1.8}$ with $\chi_{r}^2$=1.03 [\cite{barret}]). Furthermore, the electron temperature value and the optical
%depth are lower
%($\textit{kT}_\textit{e}$=17.3$^{+0.9}_{-1.2}$) with respect to the values
%already obteined by \cite{barret}. .\\
%
The ISGRI  high energy sensitivity provides a better
evaluation and error constraint of the comptonization parameters
($\textit{kT}_\textit{e}$=17.5$^{+1.7}_{-1.4}$ and $\tau$=2.3$^{+0.2}_{-0.2}$
with $\chi_{r}^2$=1.03) compared to the ones obtained by BeppoSAX
 in 2000 ($kT_\textit{e}$=36.0$^{+78.0}_{-9.0}$ keV,
$\tau$=3.0$^{+0.7}_{-1.8}$ with $\chi_{r}^2$=1.03 [\cite{barret}]).
%
%%��������������CURVE LUCE BURSTS
\begin{figure*}[ht]
 \centering
 \includegraphics[height=8cm,width=5.5cm, angle=-90]{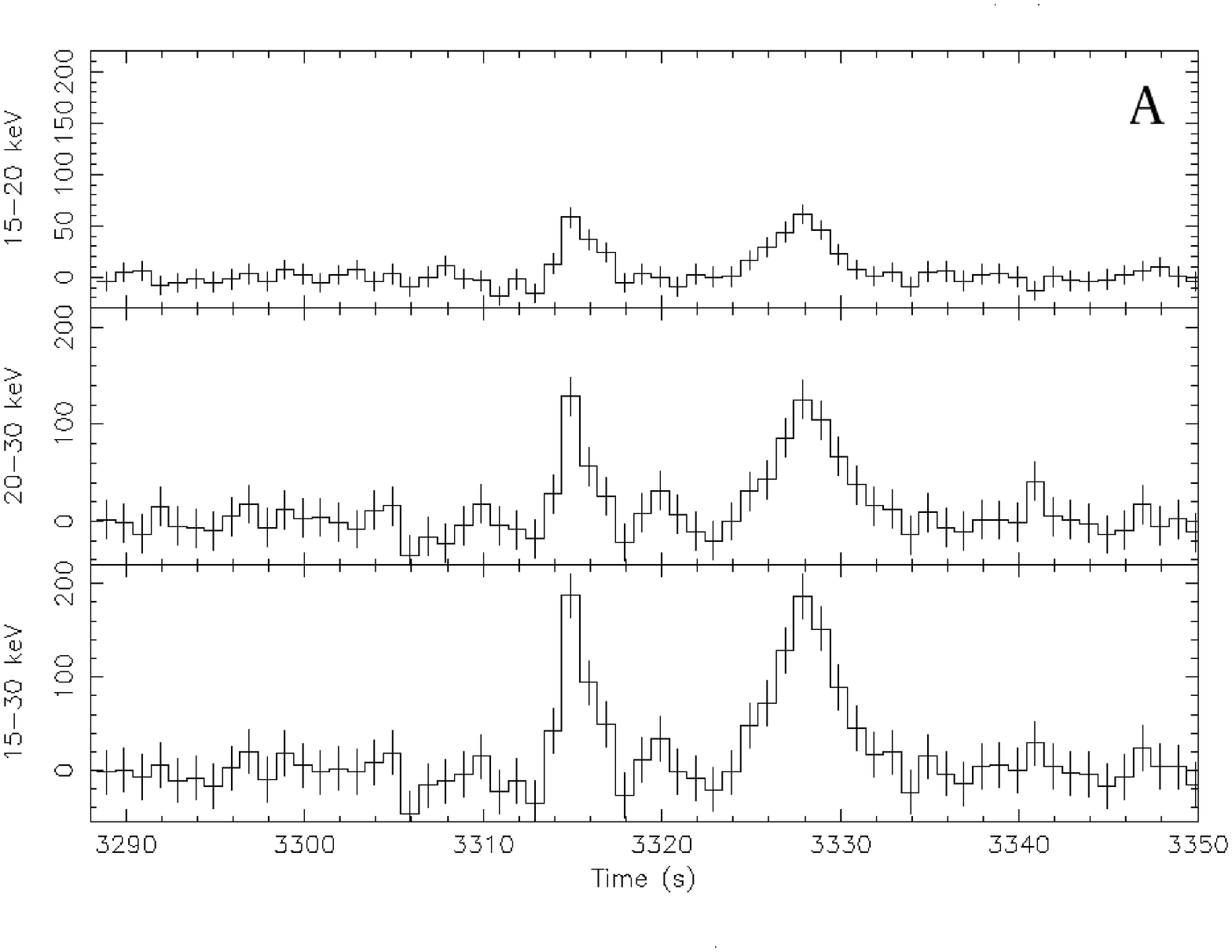}
 \includegraphics[height=7.9cm,width=5.5cm, angle=-90]{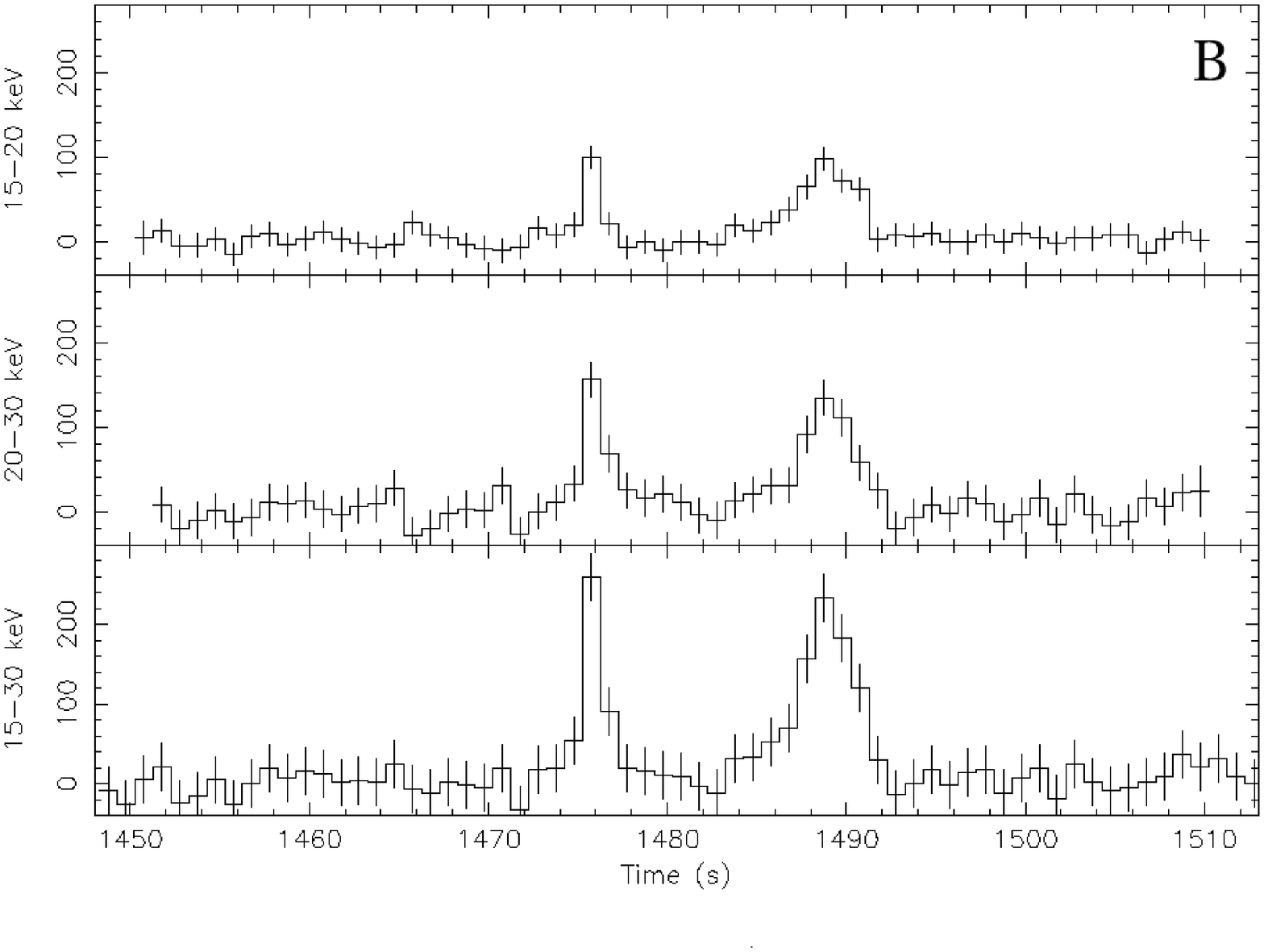}
  \includegraphics[height=7.9cm,width=5.5cm, angle=-90]{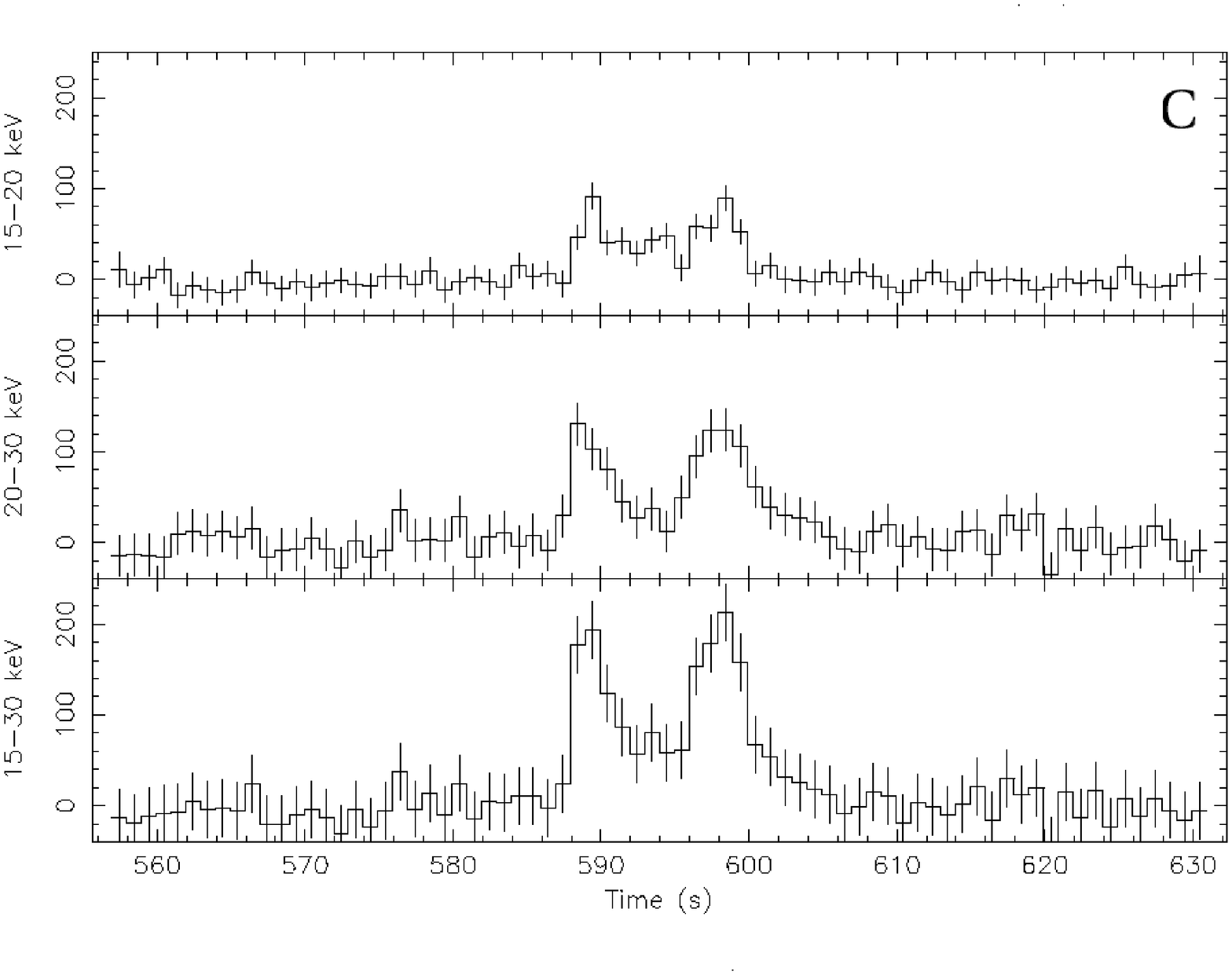}
   \includegraphics[height=7.9cm,width=5.5cm, angle=-90]{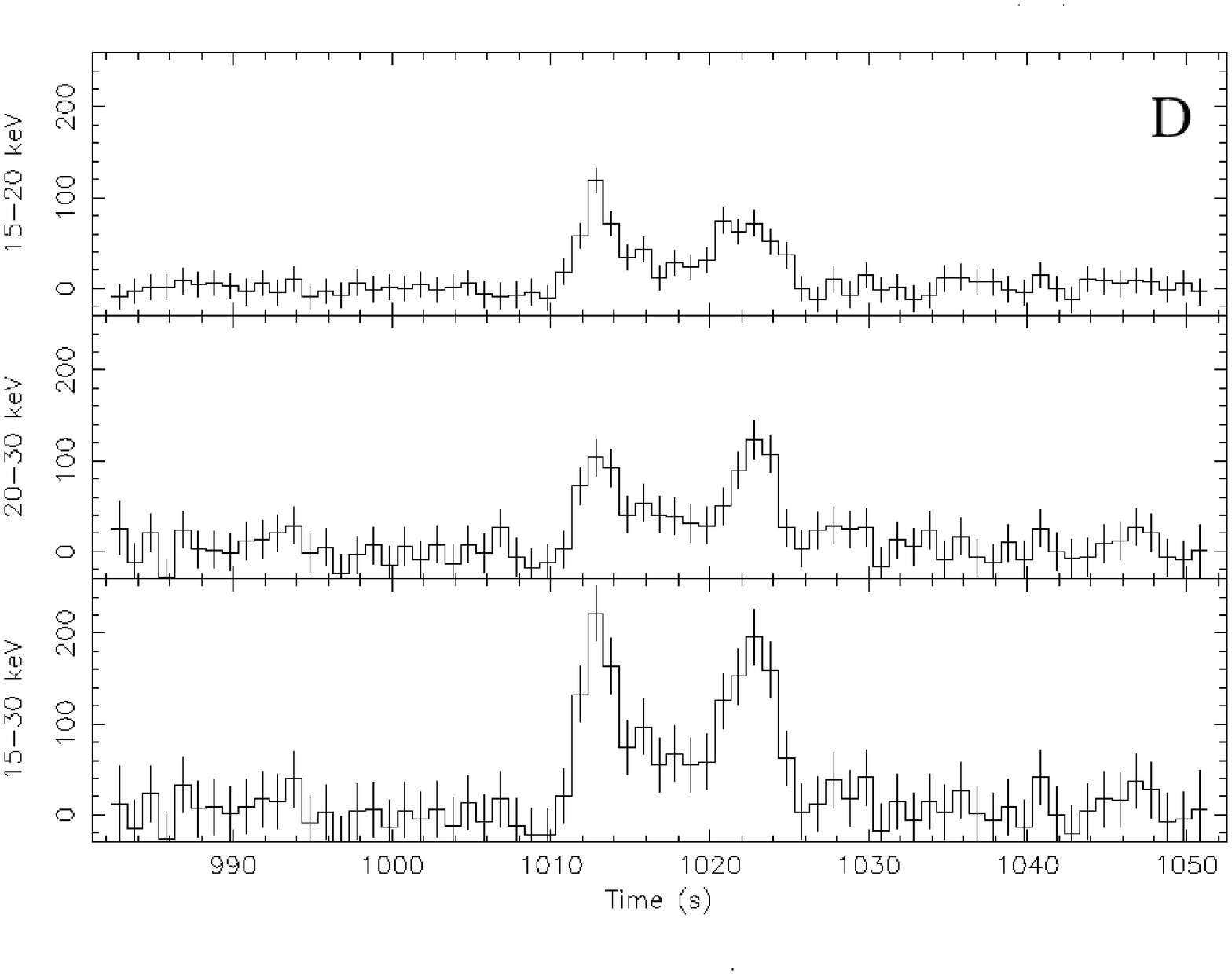}
\caption{IBIS/ISGRI light curves (counts $s^{-1}$ vs time) of the 4 bursts of 4U1812-12 in the 15-20, 20-30, 15-30 keV energy bands, with temporal bin of 1 second.}
\label{cluceburst}
 \end{figure*}
%_________________________________________$\chi_{r}^2$______________________________________________________________________>
%%
%******************************** TABELLA con i valori spettrali BURST
\begin{table*}[ht]
\begin{center}
\caption{Parameters of the burst peaks. The spectral model
      applied is the blackbody model (bbodyrad).}
\label{tabspe_bursts}
\begin{tabular}{l|cc|cc|cc|cc}
%\hline
%\cline{1-3}
%\multicolumn{1}{l}{Spectral Model: COMPTT}\\
\hline
\hline
        & \multicolumn{2}{c|}{\textbf{Burst A}} & \multicolumn{2}{c|}{\textbf{Burst B}}      & \multicolumn{2}{c|}{\textbf{Burst C}} & \multicolumn{2}{c}{\textbf{Burst D}}                           \\
        &\multicolumn{2}{c|}{2003-04-25 UT 10:54:28} & \multicolumn{2}{c|}{2003-09-06 UT 00:23:49}      & \multicolumn{2}{c|}{2003-09-27 UT 16:08:52} & \multicolumn{2}{c}{2004-10-04 UT 03:15:53} \\
\cline{2-9}
  Parameters       & 1\tiny{$^{st}$} peak & 2\tiny{$^{nd}$} peak & 1\tiny{$^{st}$} peak & 2\tiny{$^{nd}$} peak    & 1\tiny{$^{st}$} peak & 2\tiny{$^{nd}$} peak  & 1\tiny{$^{st}$} peak & 2\tiny{$^{nd}$} peak                            \\
\hline
\hline
$kT_{bb}$(keV)$^{\mathrm{a}}$ & 2.7$^{+0.6}_{-0.5}$ & 3.1$^{+0.5}_{-0.4}$
&2.7$^{+0.6}_{-0.5}$ &2.2$^{+0.4}_{-0.3}$ &  2.3$^{+0.5}_{-0.4}$ &
2.7$^{+0.4}_{-0.3}$ &1.9$^{+0.4}_{-0.3}$ &2.3$^{+0.3}_{-0.3}$  \\
%\hline
%norm$^{\mathrm{a}}$ & 0.8$^{+1.1}_{-0.4}$ & 0.5$^{+0.3}_{-0.2}$ &
%0.8$^{+1.3}_{-0.4}$ & 3.0$^{+4.6}_{-1.5}$  & 2.0$^{+4.0}_{-1.1}$ &
%0.6$^{+0.5}_{-0.2}$ & 9.3$\pm6.0$ & 2.8$^{+3.2}_{-1.3}$ \\
\hline
R$_{\textrm{km}}$/d$_{10\hspace{3pt}\textrm{kpc}}$ & 10.5$\pm$5.1 &7.0$\pm$2.5 &10.9$\pm$5.8 &33.1$\pm$10.7 &22.1$\pm$10.4 &10.3$\pm$3.7 &76.3$\pm$26.8
&28.3$\pm$9.8 \\
\hline
 $\chi_{r}^2$  &1.6 &1.2 &1.0 &1.3 &1.4 &1.3 &1.9 &1.8 \\
\hline
duration (s) & 4  & 9  & 5  & 8 & 5 & 9 & 4 & 6 \\
\hline
 Counts $s^{-1}$ & 116 $\pm18$   &    104$\pm12$ & 64$\pm17$ & 153$\pm14$             & 143$\pm19$            &  129$\pm15$       &   177$\pm21$      &  176$\pm17$         \\
\hline
% Flux$^($\footnote{$\times\cdot$10$^{-8}$ ergs cm$^{-2}$ s$^{-1}$}$^)$$_{15-30\hspace{2pt}\textrm{keV}}$ & 1.18  & 1.03  & 1.12 & 2.05 & 1.86  & 1.27  & 3.08 & 2.28\\
 Flux$^{\mathrm{a}}_{15-30\hspace{2pt}\textrm{keV}}$ & 1.18  & 1.03  & 1.12 & 2.05 & 1.86  & 1.35  & 3.08 & 2.28\\
%15-30 keV & &  & & & & & & \\
%& &  & & \\
\hline
\end{tabular}
\end{center}
\begin{list}{}{}
%\item[$^{\mathrm{a}}$]Parameter value of a blackbody model. Norm is equal to L$_{39}$/d$_{10\hspace{3pt}\textrm{kpc}}$.
%\item[$^{\mathrm{b}}$]Normalization value of the blackbodyrad model.
\item[$^{\mathrm{a}}$]In unit of 10$^{-8}$ ergs cm$^{-2}$ s$^{-1}$.
\end{list}
%\caption{\small{\emph{Parameter. $^{(1)}$ In unit of 10$^{-8}$ ergs cm$^{-2}$ s$^{-1}$}}}
%\label{tabspe_bursts}
\end{table*}
%*****************************************fine2
%
%_________________________________________________________________________________________________________________<
%%%%\clearpage
The 20-200 keV flux value, derived from ISGRI,  is 4.2$\times$10$^{-10}$ ergs cm$^{-2}$ s$^{-1}$, and the 1-200 keV flux value, combining LECS MECS and ISGRI,  is 9.1$\times$10$^{-10}$ ergs
cm$^{-2}$ s$^{-1}$.
Assuming a source distance of 4.1 kpc [\cite{cocchi}], the bolometric
luminosity is 2$\times$10$^{36}$ ergs s$^{-1}$, that corresponds to
L/L$_{Edd}$$\sim$ 0.01 (for a neutron star mass M=1.4M$_{\odot}$).\\
We have also analysed the recent publicly available RXTE/PCA spectrum data\footnote{downloaded from the on line archive of HEASARC http://heasarc.gsfc.nasa.gov/docs/archive.html}
of a 4U 1812-12 observation in July 15, 2004 to check for a possible source spectral state change. The analysis of this data shows the flux in agreement with  the SAX/IBIS one and the spectrum has similar parameters and $\chi_{r}^2$  value.
\section{Burst emission results}
\begin{figure}[ht]
\centering
\includegraphics[height=9.1cm,width=9.6cm,angle=90]{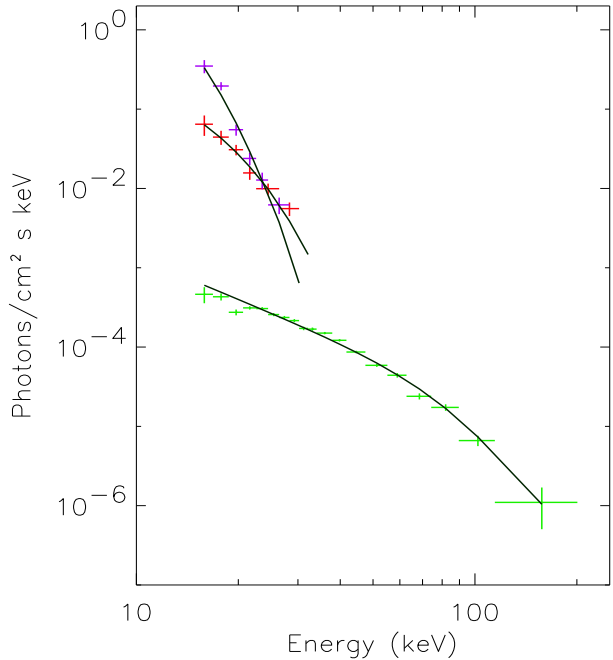}
%\includegraphics[height=6.8cm,width=5cm,angle=-90]{A_p2.ps}
 % \includegraphics[height=6.8cm,width=5cm, angle=-90]{B_p1.ps}
 %  \includegraphics[height=6.8cm,width=5cm, angle=-90]{B_p2.ps}
%\end{figure}
%\begin{figure}
% \includegraphics[height=6.8cm,width=5cm, angle=-90]{C_p1.ps}
% \includegraphics[height=6.8cm,width=5cm, angle=-90]{C_p2.ps}
 % \includegraphics[height=6.8cm,width=5cm, angle=-90]{D_p1.ps}
 %  \includegraphics[height=6.8cm,width=5cm, angle=-90]{D_p2.ps}
%
\caption{IBIS/ISGRI spectrum of the persistent emission compared with spectrum
  of the
  most and the least intense bursts. The solid lines describe the spectral models: blackbody for the
  bursts and comptonization for continuum emission.}
\label{burst-persistent}
 \end{figure}

% curve di luce
Four bursts were detected during the INTEGRAL GCDE campaign; the light curves (with a temporal bin of 1 second) for two energy bands (15-20 and
20-30 keV) and the whole band 15-30 keV are shown
in Figure \ref{cluceburst}. Above 30 keV there is no detection of
the burst emission. The upper limit of the persistent emission up to 30 keV during the bursts is about 20 counts $s^{-1}$ .\\
 The burst profiles are clearly double peaked and in each case the first pulse shows a
fast rise ($\sim$1 s) shape.
 The burst fluxes reach up to 2
Crab in the 15-30 keV band.
In Table \ref{tabspe_bursts} we report the characteristic
parameters of the 4 bursts (burst A, B, C, D taken in chronological order).\\
The burst
avarage spectra are affected (in the ISGRI band) by ``switch off'' during
the expansion so we have extracted a separate spectrum from each burst peak. We have fitted them with a black
body model and the best fit parameters are reported in Table \ref{tabspe_bursts}.\\
%%%%%%%%%%%%%%%%%%%%%%%%%%%%%
% They are well fitted with a black body
%model and the spectral parameters for
%each bursts peaks is reported in table \ref{tabspe_bursts}.
%In order to
%have an indication of the radius of the emitting region, we also reported the normalization value of blackbodyrad model, even if the error
%bars are large.
%%%%%%%%%%%%%%%%%%%%%%%%%%%%%
%Figure \ref{imaburst} shows the 15-30 keV image of the most energetic burst (D). The
%elapsed time is 14 seconds and the
%source was revealed at a level of 18.8 $\sigma$.\\
A comparison between the burst and persistent emission spectra is shown in Figure \ref{burst-persistent}.
%\begin{figure}[ht]
% \centering
% \includegraphics[height=6.5cm,width=6.5cm]{imaburst.ps}
%\caption{IBIS/ISGRI image in the 15-30 keV band of the most energetic burst
%  (burst D) with a duration of 14 seconds.}
%\label{imaburst}
%
% \end{figure}
%
%%\vspace{1.0cm}
% persistent emission
%The IBIS/INTEGRAL observation allowed to follow the high energy behaviour of 4U
%1812-12 and confirm the source to be a high energy emitter up to 200 keV.\\
%
 \section{Discussion and Conclusions}
The IBIS/ISGRI observation has allowed us to follow the high energy behaviour of 4U
1812-12.
During the long monitoring period, the light curves show that the persistent
source flux is almost constant and with the same behaviour in the 20-40, 40-60, 60-80 and 80-200 keV energy bands.\\
%We have analised the broad band energy spectrum adding the long exposure (165 ks) ISGRI spectrum with the not conteporary %PCA spectrum permitted by the persistent and not variable behaviour of the source.
Combining  the ISGRI spectrum with the non-contemporaneous LECS and MECS spectra, taken when the source was in the same spectral state, we obtain a broad band energy spectrum well represented by a soft thermal component (with a temperature of 0.6 keV) plus a Compton emission with an electron temperature, $\textit{kT}_\textit{e}$,  of $\sim$ 18 keV and an optical depth of the plasma, $\tau$, of  $\sim$ 2.3. The 1-200 keV luminosity is
2$\times$10$^{36}$ ergs s$^{-1}$. These results are in agreement with the
high energy behaviour of an atoll source in the so called Low-Hard state
[\cite{disalvo}]. Then, we confirm that 4U 1812-12 spends most of its time in this hard state (as is also shown by results from recent 2004 RXTE/PCA data) as well as the presence of a hard emission extending up to 200 keV.\\
%---------------------------origine fisica delle componenti
The hard component is due to the comptonization of soft photons in a hot
region between the neutron star and the accretion disk. The soft component can
be represented by a single-temperature black body model or alternatively by a
multicolor disk black body, so we cannot determine if
the origin of this emission is from the neutron star surface or from the accretion disk.\\
%---------------------------origine fisica delle componenti
% bursts
The 4U 1812-12 IBIS data also allowed us to study the bursting
activity, even if most of the burst emission is well below the ISGRI low energy
theshold (15 keV). In fact, because of the relatively high bursts fluxes and
the well-defined instrument response, we can still derive the spectral parameters of the bursts.\\
%In fact, the intensity of the bursts is very high resulting
%in a large amount of photons detected; this, coupled with the well known
%instrument normalization, allow a good fit of the spectral parameters of the bursts.
%Only the emitting region radius  value  don't properly converge  .\\
Type I bursts are due to thermonuclear flashes on to neutron star
surface. The spectral softening, typical of this type of burst, is not
evident because of the high energy band (15-30 keV) in which we
have observed the burst events, but the type I nature of the bursts is very
well established by previous measurements at lower energy [\cite{cocchi}].\\
The energy spectra of the detected bursts are satisfactorily described by a
blackbody model whose temperature changes in the range $kT$=1.9--3.1 keV, according to the thermal nature of these
events. This range of temperature is compatible with the temperature
of other bursts previously detected from 4U 1812-12.\\
%\\
The observed high energy emission of the 4 bursts has a double peaked profile, typical of
Eddington-limited events [\cite{xrayburst}].
The burst profiles are similar in the 15-20 and 20-30 keV ranges, and the
duration of the second peak is always longer than the first one as expected for 
double peaked bursts [\cite{xrayburst}].
Decay times and energy release are typical of the helium burning regime [\cite{stroh}]. No
evidence for different types of bursts such as superbursts [\cite{kuu}] has been
observed in our data set.
%
% !!!!!!!!!! AGGIUNGERE QUI  LE PROSPETTIVE PER IL FUTURO!!!!!!!!!!!!!!
%
\begin{acknowledgements}
This research has made use of data obtained through the INTEGRAL Science Data
Centre (ISDC), Versoix, Switzerland. Authors thanks M. Federici (IASF/Rome)
for the continuous effort to update the INTEGRAL archive in Rome.
A.T. thanks also G. De Cesare and L. Natalucci for their scientific and
data analysis support. A special thanks to the anonymous referee for useful suggestions.
This work
has been supported by Italian Space Agency by the grant I/R/046/04.
\end{acknowledgements}

%\newpage

\vspace{1.0cm}

\end{document}